\documentclass[12pt]{iopart}
\usepackage{graphicx}
\usepackage{iopams}
\begin{document}
\title[Boundary conditions at spatial infinity]
{Boundary conditions at spatial infinity for fields in Casimir calculations}

\author{V V Nesterenko}

\address{Bogoliubov Laboratory of Theoretical Physics, Joint Institute for Nuclear Research,
Dubna 141 980, Russia}
\ead{nestr@theor.jinr.ru}

\begin{abstract}
The importance of imposing proper boundary conditions for fields
at spatial infinity in the Casimir calculations is elucidated.
\end{abstract}
\section{Introduction}
 The starting formula in calculations of the Casimir energy
$E_0=(1/2)\sum \omega_n$ requires the knowledge of the spectrum of
the quantum field model under consideration \cite{RNC}. For this
aim the boundary conditions at the spatial infinity should be
imposed on the field  in the case of an unbounded configuration
space. Making here use of the {\it radiation condition} leads, as
a rule, to a {\it complex discrete energy spectrum} (open systems
with energy losses due to the outgoing waves). In this situation a
more admissible is the condition utilized in formulation of  the
scattering problem, i.e., at the spatial infinity the filed should
be {\it a sum of incoming and outgoing waves}. The spectrum of the
same Hamiltonian with scattering conditions is nonnegative but
continuous,  and the respective natural modes are not squared
integrable. In order to define a correct integration over such a
spectrum one has to introduce {\it the spectral density}, which is
expressed in terms of the Jost function of the relevant scattering
problem. The frequency equation, which is derived by making use of
the radiation condition, is related,  in a direct way, with the
Jost function. As a result the final formulas for the spectral
sums and spectral integrals turn out to be identical in the end.
Thus it is proposed a rigorous justification of the results of
Casimir calculations obtained by employment of the radiation
condition and frequency equations with discrete set of complex
roots.
\section{Complex frequencies and quasi-normal modes
in unbounded oscillating regions}

Here we show in the general case in what way complex frequencies
and quasi-normal modes appear when considering harmonic
oscillations in unbounded regions. Let a closed smooth surface $S$
divides the $d$-dimensional Euclidean space $\mathbb{R}^d$ into a
compact internal region $D_{\rm in}$ and noncompact external
region $D_{\rm ex}$. We consider here a simple scalar wave
equation
\begin{equation}
\label{2-1} \left ( \Delta -\frac{1}{c^2} \frac{\partial
^2}{\partial t^2} \right )u(t, {\textbf x})=0\,{,}
\end{equation}
where $c$ is the velocity of oscillation propagation and $\Delta $
is the Laplace operator in $\mathbb{R}^d$. For harmonic
oscillations
\begin{equation} \label{2-2} u(t,{\textbf x})=e^{-i\omega
t}u({\textbf x}) \end{equation} the wave equation (\ref{2-1}) is
reduced to the Helmholtz equation
\begin{equation} \label{2-3}
(\Delta +k^2)\,
 u(\textbf{x}) =0, \quad
k=\frac{\omega}{c}\,{.} \end{equation}

  The oscillations in the internal region $D_{\rm in}$ are described by
an infinite countable set of normal modes
\begin{equation}
 \label{2-4}
u_n(t,\textbf{x})=e^{-i\omega _n t}u_n(\textbf{x}), \quad
n=1,2,\ldots .
 \end{equation}
The spatial form of the normal modes (the functions
$u_n(\textbf{x})$) is determined by the boundary conditions which
are imposed upon the function $u(\textbf{x})$ on the internal side
of the surface $S$. These conditions should fit the physical
content of the problem under study. The set of normal modes is a
complete one. Hence any solution of  (\ref{2-3}) obeying relevant
boundary conditions can be expanded in terms of the normal
modes~$u_n(\textbf{x})$.

When considering the oscillations in the external domain $D_{\rm
ex }$ one imposes, in addition to the conditions on the compact
surface $S$, a special requirement concerning the behavior of the
function $u(\textbf{x})$ at large $r\equiv |\textbf{x}|$. Usually
for this purpose the radiation boundary conditions, proposed by
Sommerfeld~\cite{Sommerfeld,FM}, are used \begin{equation}
\label{2-5} \lim_{r\to \infty} r^{\frac{d-1}{2}}u(r)=
\textrm{const}\,{,}\qquad  \lim_{r\to\infty}r^{\frac{d-1}{2}}\left
( \frac{\partial u}{\partial r} -\rmi ku\right )= 0\,{.}
\end{equation}
For real values of the wave vector $k$ (for real frequencies
$\omega$) the solution to (\ref{2-3}), which obeys the radiation
conditions (\ref{2-5}) and reasonable boundary condition on a
compact surface $S$, identically vanishes.

If we remove the requirement of reality of the wave vector $k$,
then the statement, formulated above, does not hold. Namely, the
wave equation (\ref{2-1}) and the Helmholtz equation (\ref{2-3})
have nonzero solutions with complex frequencies, these solutions
obeying the radiation boundary conditions (\ref{2-5}) and a common
condition on a closed compact surface $S$ (for instance, Dirichlet
or Neumann conditions).

As a very simple and clear example we consider the oscillations of
electromagnetic field outside a perfectly conducting sphere of
radius $a$. In this case the electric and magnetic fields are
expressed in terms of two scalar functions $f_{kl}^{\rm{TE}}(r)$
and $f_{kl}^{\rm{TM}}(r)$ (Debye potentials) which are the radial
parts of the solutions to the scalar wave equation (\ref{2-1}).
Outside the perfectly conducting sphere placed in vacuum the
solution to the Helmholtz equation (\ref{2-3}) obeying the
radiation conditions (\ref{2-5}) has the form $(d=3)$
\begin{equation}
\label{2-6} f_{kl}(r)=C\, h^{(1)}_l\left ( \frac{\omega}{c}\,r
\right ){,}\quad r>a\,{,}
\end{equation}
where $ h^{(1)}_l(z)$ is the spherical Hankel function of the
first kind \cite{AS}. At the surface of perfectly conducting
sphere the tangential component of the electric field should
vanish. This leads to the following frequency equation for
TE-modes
\begin{equation}
\label{2-7} h^{(1)}_l\left ( \frac{\omega}{c}\,a \right
)=0{,}\quad l\geq 1
\end{equation}
and for TM-modes
\begin{equation}
\label{2-8} \frac{\rmd}{\rmd r}\left (h^{(1)}_l\left (
\frac{\omega}{c}\,r \right ) \right )=0, \quad r=a,\quad l\geq
1\,{.}
\end{equation}
The spherical Hankel function $h^{(1)}_l(z)$ is $\rme ^{\rmi z}$
multiplied by the polynomial in $1/z$  of a finite order. Hence
frequency equations (\ref{2-7}) and (\ref{2-8}) have a finite
number of roots, which are in the general case complex numbers.
For $l=1$ (the lowest oscillations) equations (\ref{2-7}) and
(\ref{2-8}) assume  the form $(z=a\,\omega/c)$
\begin{eqnarray}
h^{(1)}_l(z)&=&-\frac{1}{z}\,\rme^{\rmi z}\left ( 1+\frac{\rmi}{z}
\right )=0\quad (\mbox{TE modes}),
 \label{2-8a} \\
\frac{\rmd }{\rmd z}\left (z\, h^{(1)}_l(z)\right
)&=&-\frac{\rmi}{z^2}\,\rme^{\rmi z}\left ( z^2+\rmi z-1 \right
)=0\quad (\mbox{TM modes})\,{.}
 \label{2-9}
\end{eqnarray}
Thus the lowest eigenfrequencies are
\begin{eqnarray}
\frac{\omega}{c}&=&-\frac{\rmi}{a}\quad (\mbox{TE modes})\,{,}
\label{2-10}
\\
\frac{\omega}{c}&=&-\frac{1}{2a}(\rmi\pm\sqrt 3)\quad
(\mbox{TM modes})\,{.} \label{2-11}
\end{eqnarray}
The complex eigenfrequencies lead to a specific time and spatial
dependence of the  respective natural modes and ultimately of the
electromagnetic fields. So, with allowance of (\ref{2-10}), we
obtain
\begin{equation}
\label{2-12} \rme^{-\rmi \omega t}f_{k1}^{\rm{TE}}(r)=-\rmi
\,C\,\frac{a}{r}\,\rme^{(r-ct)/a}\left ( 1-\frac{a}{r} \right ),
\quad r\geq a\,{.}
\end{equation}

Thus, the eigenfunctions are exponentially going down in time and
exponentially going up when $r$ increases. Such time and spatial
behaviour is typical for eigenfunctions describing oscillations in
external unbounded regions, the physical content and details of
oscillation process  being irrelevant. The eigenfunctions
corresponding to complex eigenvalues are called {\it quasi-normal
modes} keeping in mind their unusual properties~\cite{Nollert}.
The physical origin of such features is obvious, in fact  we are
dealing here with {\it open systems} in which the energy  can be
radiated to infinity. Therefore in open systems field cannot be in
stationary state.

 Quasi-normal modes do not obey the standard completeness
condition and the notion of norm cannot be defined for them
\cite{Nollert}. Therefore these eigenfunctions cannot be used for
expansion of the classical field with the aim to quantize it and
to introduce the relevant Fock operators. Thus the radiation
conditions are not appropriate for the Casimir calculations.

 The same situation, with regard to quasi-normal modes, takes
place when we consider the oscillations of compound media. In this
case in both the regions $D_{\rm in}$ and $D_{\rm ex}$ the wave
equations are defined
\begin{eqnarray}
\left ( \Delta -\frac{1}{c_{\rm in}^2}\frac{\partial }{\partial
t^2} \right ) u_{\rm in}(t,\mathbf{x}) &=& 0, \quad \mathbf{x}\in
D_{\rm in}\,{,} \label{2-14}\\ \left ( \Delta -\frac{1}{c_{\rm
ex}^2}\frac{\partial }{\partial t^2} \right ) u_{\rm ex
}(t,\mathbf{x}) &=& 0, \quad \mathbf{x}\in D_{\rm ex} \label{2-15}
\end{eqnarray}
with the matching conditions at the interface $S$, for example, of
the following kind
\begin{eqnarray}  u_{\rm in}(t,\mathbf{x})=u_{\rm ex}(t,
\mathbf{x}){,} \label{2-16}
\\
\lambda_{\rm in}\frac{\partial u_{\rm in }(t,\mathbf{x})}{\partial
n_{\rm in }(\mathbf{x})}=\lambda_{\rm ex}\frac{\partial u_{\rm
ex}(t,\mathbf{x})}{\partial n_{\rm ex }(\mathbf{x})}, \quad
\mathbf{x}\in S\,{,} \label{2-17}
\end{eqnarray}
where $n_{\rm in}(\textbf{x})$ and $n_{\rm ex}(\textbf{x})$ are
normals to the surface $S$ at the point $\textbf{x}$ for the
regions $D_{\rm in}$ and $D_{\rm ex}$, respectively. The
parameters $c_{\rm in}$, $c_{\rm ex}$, $\lambda _{\rm in}$, and
$\lambda _{\rm ex}$ specify the material characteristics of the
media. At the spatial infinity the solution $u_{\rm ex}(t,
\mathbf{x})$ should satisfy the radiation conditions (\ref{2-5}).
For real $k$ we again have only zero solution in this problem,
both functions $u_{\rm in}(t, \mathbf{x})$ and $u_{\rm ex}(t,
\mathbf{x})$ vanishing. However, the wave equations (\ref{2-14})
and (\ref{2-15}) have nonzero solutions with complex frequencies,
i.e.\ quasi-normal modes, which satisfy the matching conditions
(\ref{2-16}) and (\ref{2-17}) at the interface $S$ and radiation
conditions (\ref{2-5}) at  spatial infinity. It is important, that
the frequencies  of oscillations in internal ($D_{\rm in}$) and
external ($D_{\rm ex}$) regions are the same. A typical example
here is the complex eigenfrequencies of a dielectric ball. This
problem has been investigated by Debye in his PhD thesis concerned
with the light pressure on a material ball \cite{Debye}.

The physical content of the radiation conditions is very clear.
They select only the oscillations with real frequencies caused by
external sources, which are situated in a compact spatial area.
From the mathematical point of view, these conditions ensure the
uniqueness of the solution of the nonhomogeneous boundary problems
in the external region $D_{\rm ex}$ or in the whole space ($D_{\rm
in}+D_{\rm ex}$) in the case of compound media.

When formulating the radiation  conditions in the text books, only
the real wave vector $k$ is considered. The possibility of
existence of quasi-normal modes with complex frequencies
satisfying the radiation conditions at spatial infinity with
complex wave vector $k$ is not mentioned usually. I am  aware only
of one literature source where the eigenfunctions with complex
frequencies are noted in this context. It is the article written
by Sommerfeld in the book~\cite{FM}.

 The standard physical method to escape appearance of complex
frequencies and quasi-normal modes is the following: the system
under study is placed in a sphere of a large radius $R$ and the
field is subjected to an additional  boundary condition on this
sphere. The initial differential operator has now a real discrete
spectrum, and the respective eigenfunctions are normalized in
$L_2$. Upon conducting the calculations the limit $R\to \infty$
should be taken.

 However, in order to justify this approach one has to prove that the
final result does not depend on the explicit form of auxiliary
boundary conditions imposed on the large sphere $r=R$. Obviously,
it is nontrivial task. Furthermore, in this approach always the
feeling is rest, that something is lost because the region $r>R$
is actually  ignored. Fortuitous, there is a rigorous way to
consider the open systems by making use of the formalism of the
scattering problem without introducing a large auxiliary sphere.

\section{Description of open systems by scattering method}
In the scattering problem the same differential operator (for
example, the Laplace operator (\ref{2-3}))  is considered, but
instead of the radiation conditions (\ref{2-5}) it is required
that the eigenfunctions $\phi_l(k,r)$ at spatial infinity $r\to
\infty$ are reduced to the sum of incoming and outgoing waves
\begin{equation}
\label{3-1}  \phi_l(k,r)\simeq -\frac{1}{2\rmi k}\left [
F_l(-k)\,\rme ^{-\rmi kr }- F_l(k)\,\rme ^{\rmi kr } \right
]{,}\quad r\to\infty \,{.}
\end{equation}
Here $F_l(k)$ is the Jost function, and spherical symmetry in the
problem under consideration is expected. The operators of the type
\begin{equation}
\label{3-2} - \Delta + V(r)
\end{equation}
have real continuous spectrum
\begin{equation}
\label{3-3} 0\leq k^2<\infty\,{,}
\end{equation}
when the potential $V(r)$ obeys known conditions at $r=0$  and at
infinity  $r\to \infty$.

Due to the continuity of the spectrum the corresponding
eigenfunctions are not normalized in  $L_2$. Making here use of
the normalization on the Dirack $\delta$-function proves to be
useful in many physical applications, however it does not allow
one to define, in a correct way, the integration over a continuous
spectrum. Indeed, let $\psi_{\mathbf{k}}(\mathbf{x})$ be the
eigenfunctions corresponding  to the spectrum (\ref{3-3})
normalized on the Dirack $\delta$-function
\begin{equation}
\label{3-4} \int
\psi^*_{\mathbf{k}}(\mathbf{x})\psi_{\mathbf{k}'}(\mathbf{x})\,\rmd^3
\mathbf{x}= \delta^{(3)}(\mathbf{k}- \mathbf{k}')\,{,}
\end{equation}
for example, those can be plane waves
\begin{equation}
\label{3-5}
\psi_{\mathbf{k}}(\mathbf{x})=\frac{1}{(2\pi)^{3/2}}\,\rme^{\rmi
\mathbf{k x}}{.}
\end{equation}
Due to its meaning, the spectral density $\rho(k)$ should be
defined in the following way
\begin{equation}
\label{3-6}
\rho(k)=\int\psi^*_{\mathbf{k}}(\mathbf{x})\psi_{\mathbf{k}}(\mathbf{x})\,\rmd^3
\mathbf{x}=\delta^{(3)}(0)=\frac{V}{(2\pi )^{3}}{,}
\end{equation}
where $V$ is the volume of the space $\mathbb{R}^3$ and obviously
$V\to \infty$. It is worth noting that the spectral density
(\ref{3-6}), considered just on the formal footing, does not
depend on the details of the dynamics, i.e.\ on the potential
$V(r)$ and corresponding boundary conditions, therefore it can not
be used in physical calculations.

  In the rigorous mathematical  theory of scattering processes  this problem
has been solved. In the most simple way the spectral density for
scattering states can be derived as a consequence of a {\it
spectral property}~\cite{OB}. Let $r(z)=(h-z)^{-1}$ and
$r_0(z)=(h_0-z)^{-1}$ be the resolvents of a complete ($h$) and
free ($h_0$) Hamiltonians. When the potential meets the relevant
conditions, then the relation, known as the spectral property,
holds
\begin{equation}
\label{3-7} 2\, {\rm Im}\,\Tr \left [ r(E+\rmi 0)-r_0(E+\rmi 0)
\right ] = \tr\,  [q(E)] \,{,}
\end{equation}
where
\begin{equation}
\label{3-8} q(E)= -\rmi S^\dag(E) \frac{\rmd}{\rmd E}\, S(E)
\end{equation}
is the two-body time delay operator and $S$ is the scattering
matrix
\begin{equation}
\label{3-8a} S(k)=\rme ^{2\rmi \delta(k) }=\frac{F(k)}{F(-k)}.
\end{equation}
Here $\delta (k) $ is the phase shift and $F(k)$ is the Jost
function.

  By making use of the spectral representation for the
resolvents and definition (\ref{3-6}) one obtains immediately
\begin{eqnarray}
\label{3-9} \Delta \rho (k)& \equiv& \rho(k) - \rho_0 (k) =
 \frac{1}{2\pi\rmi}\,\frac{\rmd }{\rmd k}\,\ln
S(k)\nonumber \\&=&\frac{1}{2\pi\rmi}\,\frac{\rmd }{\rmd k}\,\ln
\frac{F(k)}{F(-k)}=\frac{1}{\pi}\,\frac{\rmd}{\rmd k}\, \delta
(k)\,{.}
\end{eqnarray}
It is worth noting that in this formula the contribution of a free
unbounded space is already subtracted. The finite expression for
$\Delta \rho (k)$ is a direct consequence  of considering complex
values of the wave vector $k$ and respective energy~$k^2$.

\section{Correct transition to imaginary frequencies in the case of unbounded domains}
 In the Casimir calculations the spectrum of quantum field systems is not known explicitly.
As a rule, it is given by the roots ($\omega_n$) of a frequency
equation
\begin{equation}
\label{4-1} f(\omega)=0
\end{equation}
with the known function $f$. In order to sum these roots the
counter integration can be used
\begin{equation}
\label{4-2} \frac{1}{2}\sum_n
\omega_n=\frac{1}{2\pi\rmi}\oint_C\frac{z}{2}\,\frac{\rmd}{\rmd z}
\ln f(z)\, \rmd z{,}
\end{equation}
where the counter $C$ encloses, counterclockwise, all the  roots
of  (\ref{4-1}).

For a compact configuration space without dissipation the roots of
(\ref{4-1}) are real and positive. The counter $C$ can be deformed
continuously into the counter $C_R$ with $R\to \infty$ (see figure
1a). Usually $f(x)=f(\sqrt{x^2})$, therefore $f(x)=f(-x)$. The
integral along the semicircle of radius $R$ vanishes when $R\to
\infty$ or it is canceled by the subtraction of the free space
contribution. As a result we obtain
\begin{equation}
\label{4-3} \frac{1}{2}\sum_n \omega_n
=-\frac{1}{\pi}\int_0^\infty \frac{y}{2} \,\frac{\rmd}{\rmd y} \ln
f(\rmi y)\,\rmd y\,{.}
\end{equation}

Let us derive the analog of (\ref{4-3}) for the open systems.
According to the  inferences of the previous section we have to
consider the following integral along the continuous spectrum
\begin{equation}
\label{4-4} \frac{1}{2}\int _0^\infty k\, \Delta \rho (k)\, \rmd
k=\frac{1}{2\pi \rmi }\int _0^\infty\frac{k}{2}\,\frac{\rmd }{\rmd
k }\ln\frac{F(k)}{F(-k)}\,\rmd k\,{,}
\end{equation}
where $F(k)$ is the Jost function in the pertinent scattering
problem. This function is analytic in the lower half plane of the
complex variable $k$. We assume that there are no bound states in
the system in question. Under this assumption the zeros of the
Jost function $F(k)$ can lay only in the upper half plane $k$
\cite{Newton,NewtonJMP}. In view of this we infer that the counter
integrals (see figure 1b)
\begin{equation}
\label{4-5}
\frac{1}{2\pi\rmi}\oint_{C^{\mp}_R}\frac{k}{2}\,\frac{\rmd}{\rmd k
}\ln F(\pm k)\, \rmd k =0
\end{equation}
vanish  because inside the counter $C^+_R \; (C^-_R)$ there are no
zeros and singularities of the function $F(-k) \; (F(k) )$. Now we
can transform in (\ref{4-4}) the integration along the real
positive axes $k$ to the integration along the imaginary
frequencies $k=\rmi y,\quad y>0$ with the result
\begin{equation}
\label{4-6} \frac{1}{2}\int_0^\infty k\,\Delta \rho (k)\,\rmd
k=-\frac{1}{\pi}\int_0^\infty\frac{y}{2}\,\frac{\rmd}{\rmd y}\ln
F(-\rm i y)\,\rmd y\,{.}
\end{equation}

\noindent
\begin{figure}[th]
\noindent \centerline{
\includegraphics[width=45mm]{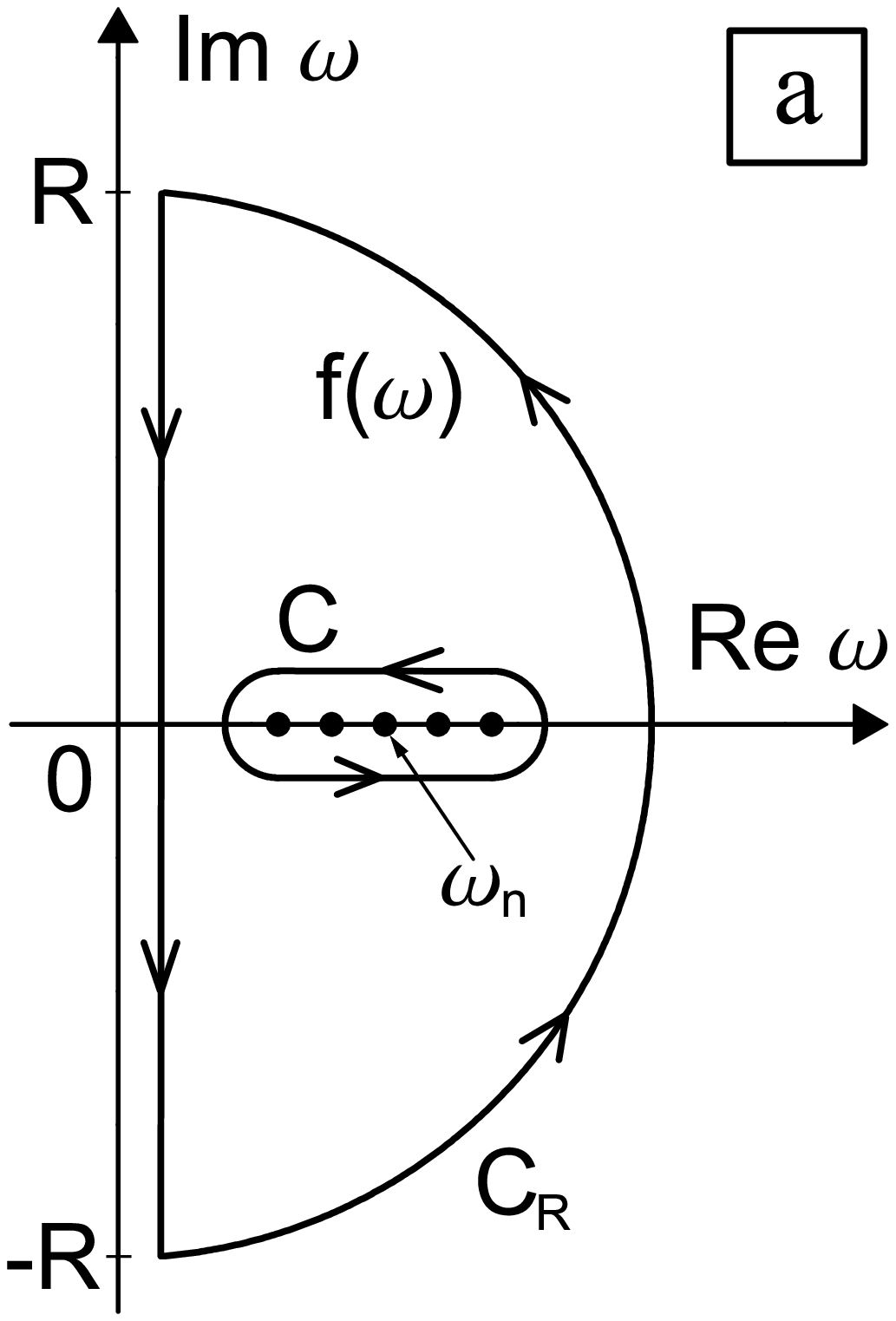}
\hspace{20mm}
\includegraphics[width=45mm]{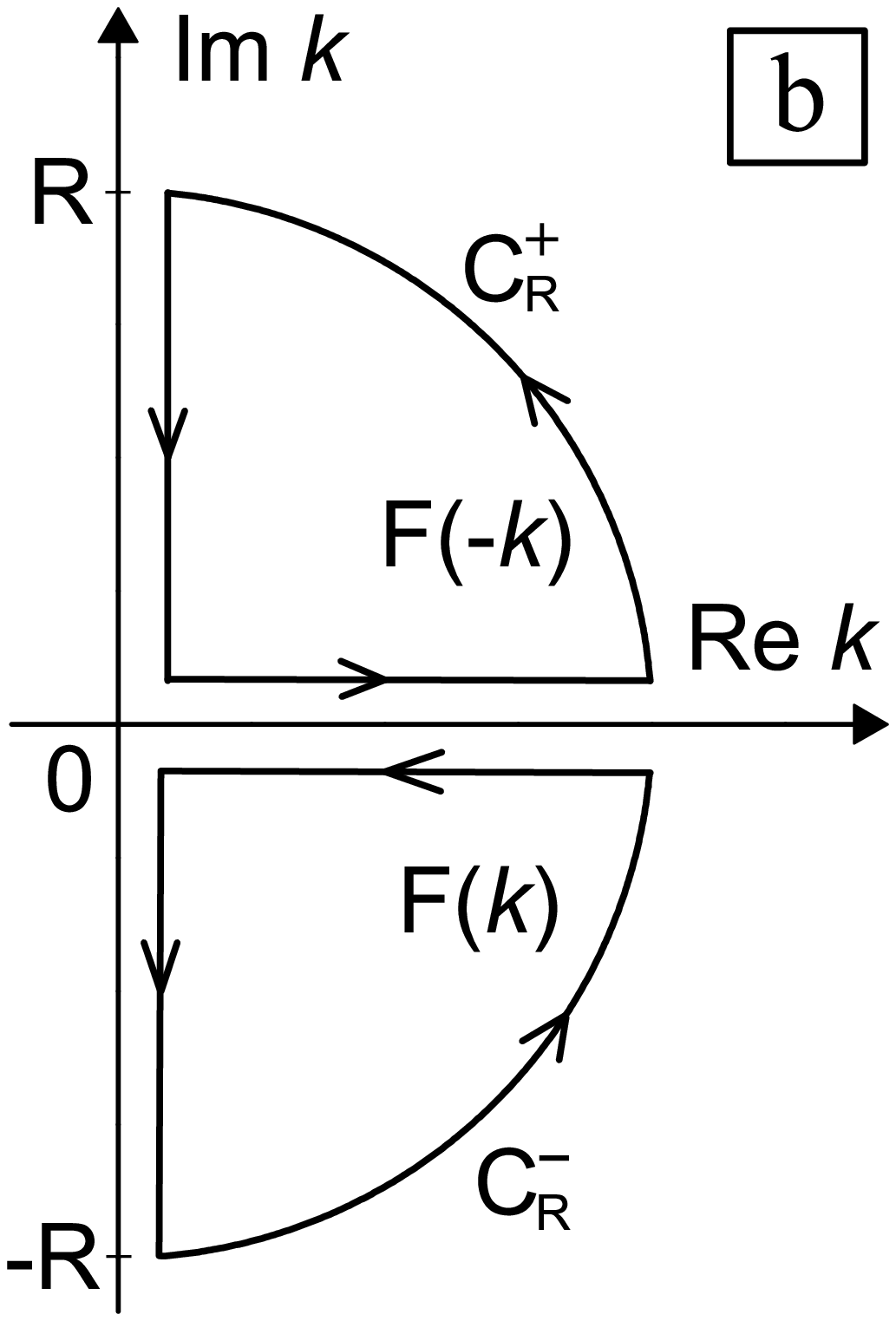}}
\caption{The contours on the complex $k$ plane which are used when
going on to the integration over the imaginary  frequencies in the
case of bounded (a) and unbounded (b) configuration spaces.}
\label{Plot:Contours2}
\end{figure}

Now we take into account the following. The function $f(\omega)$
in frequency equation (\ref{4-1}), which is derived by making use
of the radiation conditions at spatial infinity, is actually the
Jost function up to irrelevant multiplier. Indeed, the solution
(\ref{3-1}) will contain only outgoing waves when
\begin{equation}
\label{4-7} F(-k)=0\,{.}
\end{equation}
For simplicity we have dropped here the index $l$. Thus we have
\begin{equation}
\label{4-8} f(\omega)=F(-k), \quad k^2=\omega^2/c^2.
\end{equation}
Now we can infer that the right-hand sides of  (\ref{4-3}) and
(\ref{4-6}) coincide. Hence, the formula (\ref{4-3}) with the
function $f$ defining the frequency equation is applicable also to
unbounded configuration spaces.

All this removes the objections \cite{BH,Hagen} against using
(\ref{4-3}) for unbounded regions and explains why it gives
correct results there~\cite{RNC}.

\section{Conclusion}
Importance of proper choosing conditions, imposed on the field at
spatial infinity when determining the physical spectrum in Casimir
calculations, is illustrated by a circular infinite dielectric
cylinder. If we confine our consideration to the oscillating
asymptotics for $r\to \infty$ the spectrum of Maxwell equations
is continuous. Only this situation has been considered in all
pertinent papers (see, for example, \cite{BP}). However from the
physical point of view are acceptable also solutions exponentially
going down when $r\to \infty$. It is these configurations of
electromagnetic field (surface waves with discrete spectrum) that
are used in dielectric wavegaudes~\cite{Jackson}. Thus the
question arises whether this branch of spectrum has been taken
into account in relevant calculations. Casimir studies, which
allow for the material properties of the boundaries, show
importance of the surface waves, or more precisely, of surface
plasmon contributions \cite{BPN} at short distances (regime
without retardation). Obviously, such bound states should be added
to the spectral density~(\ref{3-9}).

 \ack It is a pleasant duty for me to  thank Professors Emilio
Elizalde and Sergei Odintsov for excellent organization of the
Workshop QFEXT'05 and for arranging my participation in this
Meeting. The partial financial support of Russian Foundation for
Basic Research (Grants No.\ 05-01-10697 and 03-01-00025) is
acknowledged. I am grateful to A V Nesterenko for preparing the
figure.


\section*{References}
\begin{thebibliography}{99}
\bibitem{RNC} Nesterenko V V, Lambiase G and Scarpetta G 2004 {\it
Rivista del Nuovo Cimento} {\bf 27} No.~6, 1--74
\bibitem{Sommerfeld} Sommerfeld A 1949 {\it Partial Differential Equations of
Physics} (Academic Press, New York)
\bibitem{FM} Frank P and Mises R 1961 {\it Die Differential- und Integralgleihungen
der Mechanik und Physik} II Physikalischer Teil (Dover
Publications, New York)
\bibitem{AS}Abramowitz M and Stegun I (eds) 1972 {\it Handbook of
Mathematical Functions} (Dover Publications, New York)
\bibitem{Nollert} Nollert H-P 1999 {\it Class.\ Quantum Grav.} {\bf 16}
R159
\bibitem{Debye} Debye P  1909 {\it Ann. Phys. (Leipzig)} {\bf 30}
57
\bibitem{OB} Osborn T A and Bolle D 1977 \JMP {\bf 18} {432}
\bibitem{Newton} Newton R G 1982 {\it Scattering Theory of Waves and Particles}
2nd~ed (Springer-Verlag, New York)
\bibitem{NewtonJMP} Newton R G 1960 \JMP {\bf 1} 319
\bibitem{BH} Bowers M E and Hagen C R 1999 \PR D {\bf 59}
025007
\bibitem{Hagen} Hagen C R 2000 \PR D {\bf 61} 065005
\bibitem{BP} Bordag M and Pirozhenko I G  2001 \PR D {64} 025019
\bibitem{Jackson}  Jackson J D 1975 {\it Classical Electrodynamics} 2nd~ed  (Wiley, New York)
\bibitem{BPN} Bordag M, Pirozhenko I G and Nesterenko V V Spectral
analysis of a flat plasma sheet model {\it Preprint}
hep-th/0508198 (to be published in \JPA)
\end{thebibliography}
\end{document}